\documentclass[fleqn,usenatbib,useAMS]{mnras}
\usepackage{amssymb,amsmath}
\usepackage{graphicx}
\usepackage{xcolor,natbib}
\usepackage{multirow}
\usepackage[T1]{fontenc}
\usepackage{ae,aecompl}
\usepackage{newtxtext, newtxmath}
\usepackage{float}
\usepackage{subfigure}
\usepackage{longtable}
\usepackage{enumitem}
\usepackage{longtable,listings}
\usepackage[flushleft]{threeparttable}
\usepackage{parcolumns}
\def\dif{\mathop{}\!\mathrm{d}}
\def\obj{SDSS J1321}

\title[3.8yr optical QPOs in SDSS J1321]{A 3.8yr optical quasi-periodic oscillations in blue quasar SDSS 
J132144+033055 through combined light curves from CSS and ZTF}
\author[Zhang X. G.]{XueGuang Zhang$^{1}$
\thanks{Contact e-mail: \href{mailto:aexueguang@qq.com}{aexueguang@qq.com}}\\
$^{1}$School of Physical Science and Technology, Guangxi University,
          No. 100, Daxue East Road, Nanning, 530004, P. R. China}

\begin{document}
\label{firstpage}
\pagerange{\pageref{firstpage}--\pageref{lastpage}}

\maketitle

\begin{abstract} %%%about 248 words
	In the manuscript, a 3.8yr optical quasi-periodic oscillations (QPOs) is reported in blue 
quasar SDSS J132144+033055 (=\obj) at $z=0.269$, based on 16.3yr-long light curve from both CSS and 
ZTF directly described by a sinusoidal function. The 3.8yr QPOs can be confirmed through the Generalized 
Lomb-Scargle periodogram with confidence level higher than 5$\sigma$, through properties of the 
phase-folded light curve and the WWZ technique. Moreover, the collected Pan-STARRS light curves well 
follow the sinusoidal function described best fitting results to the CSS and ZTF light curves. The 
optical QPOs strongly indicate a central binary black hole (BBH) system in \obj, with 
expected space separation smaller than 0.018pc, through the estimated upper limit of total BH mass 
$3.3\times10^9{\rm M_\odot}$ through the correlation between BH mass and continuum luminosity.  
Meanwhile, we check disk precession applied to explain the optical QPOs. However, under the disk 
precession assumption, the determined optical emission regions from central BH have sizes about 
$37{\rm R_G}$ similar as the sizes $35{\rm R_G}$ of the expected NUV emission regions through the 
correlation between disk size and BH mass, indicating the disk precession is not preferred. And due 
to undetected radio emissions, jet precession can be ruled out. Furthermore, only 0.1\% probability 
can determined as the QPOs mis-detected through CAR process randomly created light curves related to 
intrinsic AGN activities, re-confirming the optical QPOs with significance level higher 
than 3$\sigma$. Therefore, combining long-term light curves from CSS and ZTF can lead to more QPOs 
candidates in the near future. 
\end{abstract}


\section{Introduction}

%%first
	Merging of galaxies is an essential process of galaxy formation and evolution \citep{cr92, 
lc93, kw93, bh96, sr98, mh01, lk04, md06, bf09, se14, rp16, rs17, bh19, mj21, yp22}, leading to common 
dual galactic core systems on scale of dozens to hundreds parsecs to supermassive binary black hole 
(BBH) systems on scale of sub-parsecs in galaxies \citep{bb80, mk10, fg19, mj22}. Different techniques 
have been applied to detect dual core systems with the two black holes getting closer due to dynamical 
friction and/or BBH systems with two black holes getting closer due to emission of gravitational waves. 
Double-peaked features of broad and/or narrow emission lines has been accepted as signs of dual core 
systems and BBH systems, as the reported results in \citet{zw04, kz08, bl09, ss09, sl10, eb12, pl12, 
cs13, le16, wg17, dv19}. Spatially resolved high quality image properties of central regions of galaxies 
have been applied to detect dual core systems and/or BBH systems, as discussed in \citet{km03, rt09, 
pv10, ne17, kw20, sv21, sb21}. More recently, \citet{zh21d} have reported the different broad Balmer 
emission line features as the sign to support a central BBH systems in SDSS J154751.94+025550.8  
with double-peaked broad H$\beta$ but single-peaked broad H$\alpha$. Moreover, apparent Optical 
Quasi-Periodic Oscillations (QPOs) signals have been well applied to detected BBH systems in galaxies.

%%%2
	QPOs with periodicities of years to more than ten years have been reported in active galaxies, 
due to jet emissions/precessions in blazars as discussed in \citet{sc18, bg19, oa20} or due to BBH 
systems as discussed in \citep{eh94, kf09, gm10, bb15, sx20}. Meanwhile, there is one special kind of 
QPOs, transient QPOs, arising from general relativistic effects (relativistic Frame Dragging method 
\citep{ref1}, discoseismology method \citep{ref2}, etc.) related to central accreting processes, such 
as the reported transient QPOs in black hole X-ray binaries in \citet{vm98, vm00, ak04, rm06, jp10, 
vi17, im20} and in several Active Galactic Nuclei (AGN) in \citet{pl93, mk06, gm08, li13, ps14, rr16, 
dl17, bs18, sm18, gt18, jd21, zh21a}. In the manuscript, rather than the transient QPOs related to 
general relativistic effects, we mainly consider the long-standing QPOs related to orbital motions of 
two BH accreting systems in BBH systems which can be directly detected in the long-term light curves.

%%%3
	In the literature, there are hundreds of QPOs related to BBH systems. 1800days optical QPOs 
have been reported in \citet{gd15a, lg18, kp19}, to support a central BBH system in the well-known 
Palomar-Green quasar PG 1302-102. Strong optical QPOs with periodicities of hundreds to 
thousands of days have been detected and reported in \citet{gd15} for a sample of 111 candidates of 
central BBH systems, through nine-years long-term variabilities. Significant QPOs with periodicities 
of a few hundred days have been reported in 50 quasars in \citet{cb16}. Moreover, BBH system expected 
540days QPOs have been reported in the quasar PSO J334.2028+01.4075\ in \citet{lg15}, 1500days QPOs 
have been reported in the Sloan Digital Sky Survey (SDSS) quasar SDSS J015910.05+010514.5  
in \citet{zb16}, 1150days QPOs have been reported in the Seyfert1.5 Mrk 915\ in \citet{ss20}, 1.2yr 
QPOs have been reported in the Mrk 231\ in \citet{ky20}, 1607days QPOs have been reported in the quasar 
SDSS J025214.67-002813.7 in \citet{lw21}. More recently, \citet{zh22} have reported 6.4yr 
optical QPOs in SDSS J075217.84+193542.2, to support a central BBH system.

%%%%4
	The BBH systems can produce expected background gravitational wave signals at nano-Hz frequencies 
probed by the Pulsar Timing Arrays (PTA) \citep{fb90, de16, re16, ar15, ve16}. However, besides the reported 
optical QPOs to support central BBH systems, false periodicities have been discussed in quasar time-domain 
surveys, such as the results in \citet{vu16} and in \citet{se18}. \citet{vu16} have shown that 
false periodicities can come from intrinsic AGN variabilities well dominated by stochastic process, and 
discussed the importance of calibrating the false positive rate of detecting optical QPOs. Moreover, 
through properties of expected gravitational wave background at nano-Hz frequencies probed by the 
PTA, \citet{se18} have shown that the null hypothesis (whereby the candidates of BBH systems are false 
positives) is preferred over the BBH hypothesis at about 2.3$\sigma$ and 3.6$\sigma$ for the BBH candidates 
in \citet{gd15} and in \citet{cb16} respectively, indicating the current candidates of BBH systems have 
some false candidates due to false QPOs detections. Therefore, it is necessary and meaningful to detect 
and report more candidates of BBH systems.

%%%5
	As what have been discussed in \citet{gd15}, time durations longer enough are necessary 
and better to detect reliable optical QPOs. Therefore, combining light curves from different sky 
surveys covering different time epochs can lead to longer time durations of light curves of objects, 
providing efficient information to detect more reliable optical QPOs. And in this manuscript, a new 
BBH candidate is reported in the blue quasar SDSS J132144+033055 (=\obj) at a redshift 0.269, due 
to detected optical QPOs through the combined long-term variabilities from the Catalina Sky Survey 
(CSS) \citep{dr09} and from the more recent Zwicky Transient Facility (ZTF) \citep{bk19, ds20}. 
The time duration of the combined light curve of \obj~ is more than 4 times longer than the detected 
periodicity, indicating the optical QPOs in \obj~ should be robust to some extent. The manuscript 
is organized as follows. Section 2 presents main results on the long-term optical variabilities 
of \obj, to report the detected optical QPOs and the method to confirm the optical QPOs not from 
central intrinsic AGN activities. Section 3 shows main results on the spectroscopic properties of 
\obj. Section 4 gives the necessary discussions on the probable central BBH system. Section 5 gives 
final summaries and conclusions. In the manuscript, the cosmological parameters have been adopted 
as $H_{0}=70{\rm km\cdot s}^{-1}{\rm Mpc}^{-1}$, $\Omega_{\Lambda}=0.7$ and $\Omega_{\rm m}=0.3$.

\begin{figure*}
\centering\includegraphics[width = 8cm,height=6cm]{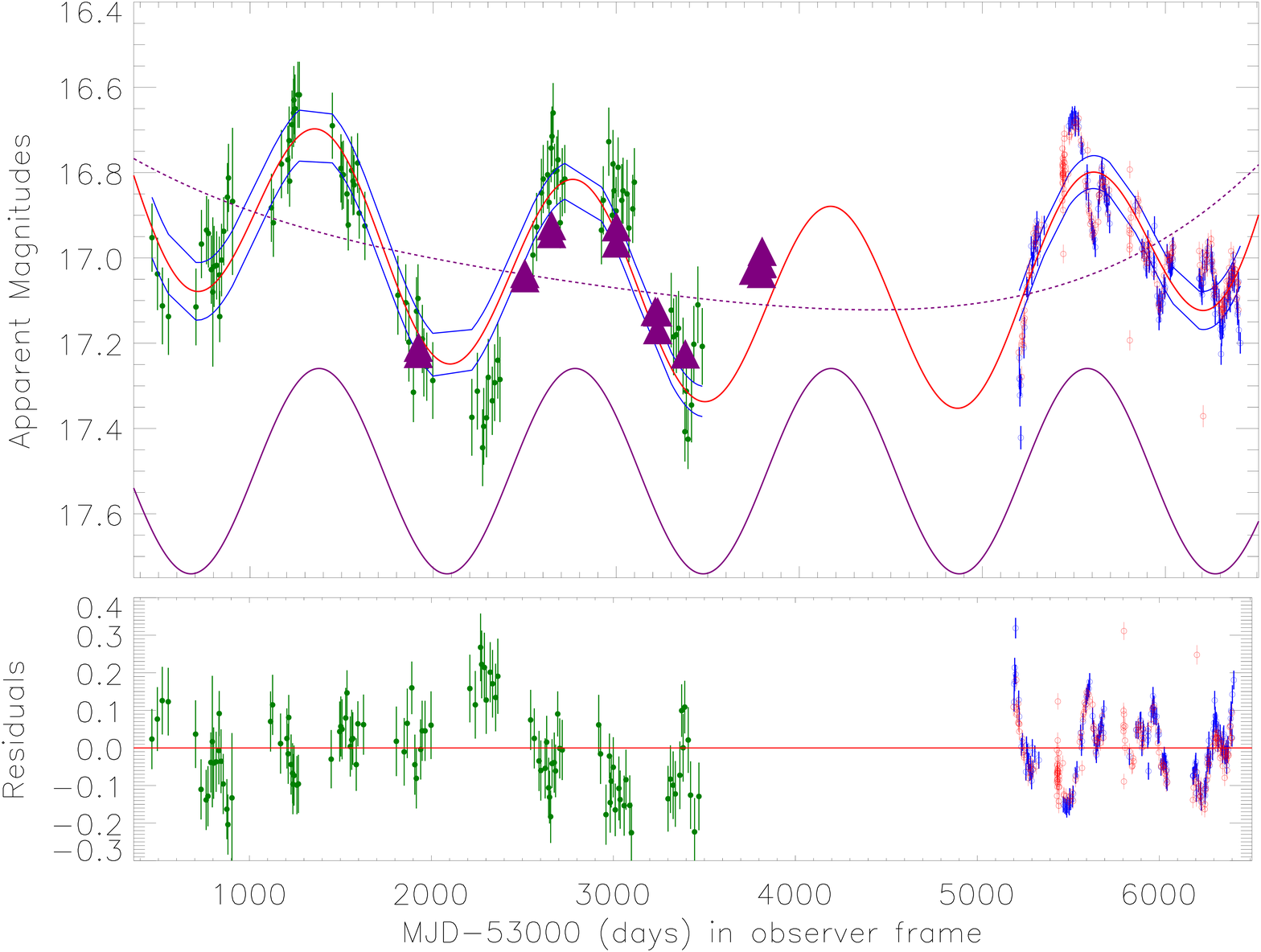}
\centering\includegraphics[width = 8cm,height=6cm]{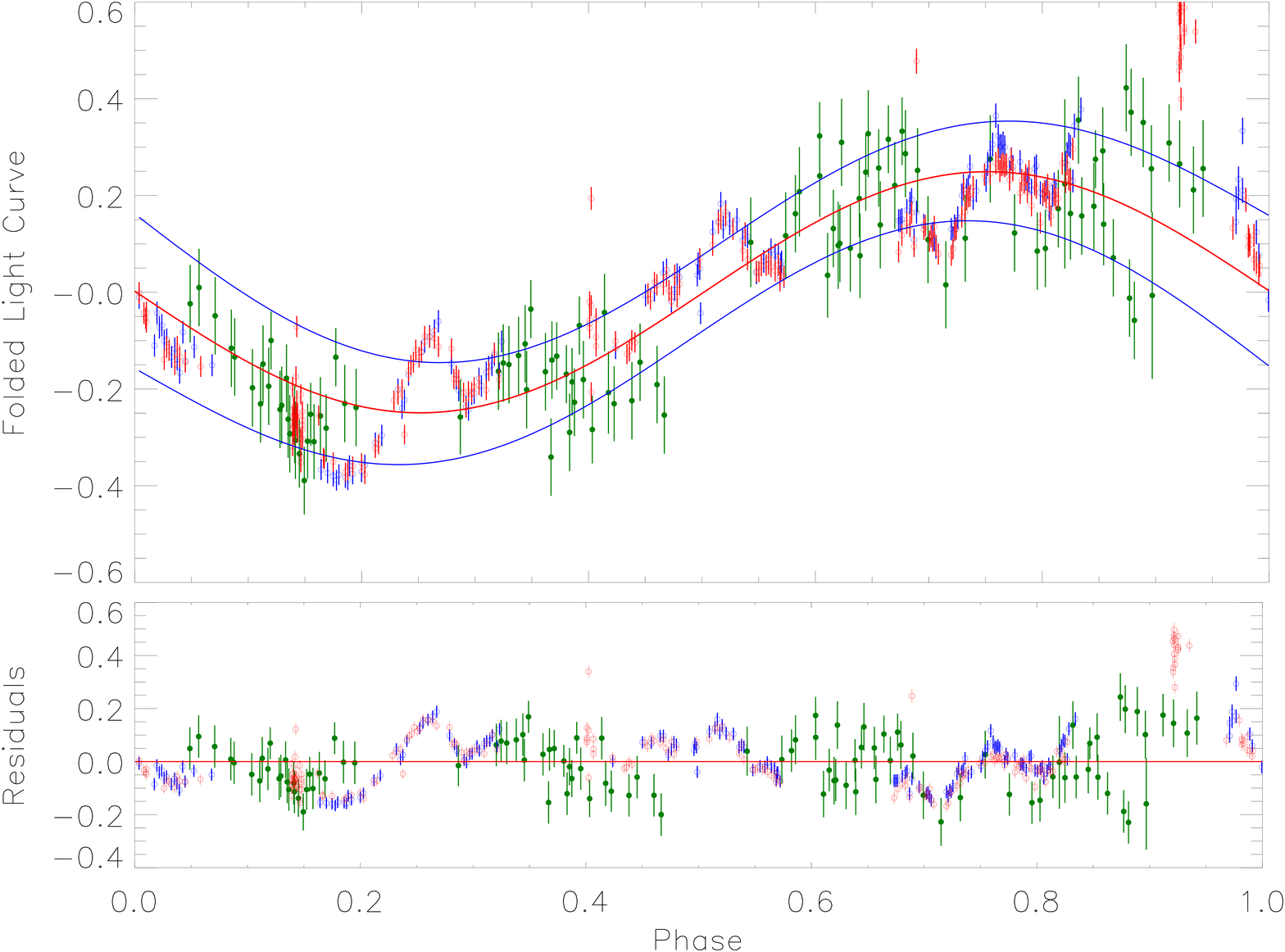}
\caption{Top left panel shows the long-term light curves from the CSS (solid circles plus error 
bars in dark green) and from the ZTF (open circles plus error bars in blue and in red for g-band 
and r-band data points, respectively) and the corresponding best fitting results (solid red line) 
by a sinusoidal function plus a four-degree polynomial function. Top right panel shows the 
corresponding phase folded light curve based on the determined periodicity of 1400days and the 
best-fitting results (solid red line) by a sinusoidal function, with symbols having the same 
meanings as those in left panel. In top left panel, solid purple triangles show the $z$-band 
light curve (with magnitudes plus 0.3) collected from Pan-STARRS, dashed purple line 
shows the determined component described by the four-degree polynomial function, solid purple line 
shows the determined component (with magnitude plus 17.5) described by the sinusoidal function.  
In each top panel, solid blue lines show the corresponding F-test determined 99.9999\% confidence 
bands to the best fitting results. Bottom panels show the corresponding residuals. In each bottom 
panel, solid red line shows Residuals~=~0.
}
\label{lmc}
\end{figure*}

\section{Long-term optical variabilities in \obj}

%%%1
\subsection{Optical QPOs in \obj}

	The CSS V-band light curve of \obj~ (RA=13:21:44, DEC=03:30:55) is collected from 
\url{http://nesssi.cacr.caltech.edu/DataRelease/} with MJD-53000 from 464 (April 2005) to 3470 
(June 2013), and the ZTF $gr$-band\footnote{There are only 21 data points in the ZTF $i$-band 
light curve of \obj, therefore, there are no considerations or discussions on properties of ZTF 
$i$-band light curve of \obj~ in the manuscript.} light curves of \obj~ are collected from 
\url{https://www.ztf.caltech.edu} with MJD-53000 from 5202 (March 2018) to 6409 (July 2021). 
Top left panel of Figure~\ref{lmc} shows the 16.3years-long photometric light curves.

%%%2
	By Levenberg-Marquardt least-squares minimization technique, a simple sinusoidal function 
plus a four-degree polynomial function are applied to determine the best descriptions to the 
16.3yr-long light curves. Here, the main objective of applications of sinusoidal function is to 
show clearer clues on optical QPOs, not to discuss physical origin of the QPOs. And the main 
objective of applications of four-degree polynomial function is to modify the probable magnitude 
difference between CSS V-band light curve and ZTF $gr$-band light curves in \obj. The best-fitting 
results with corresponding $\chi^2/dof~\sim2308.16/247\sim~9.3$ (the summed squared residuals divided 
by degree of freedom) are shown as solid red line in top left panel of Figure~\ref{lmc} by the formula
\begin{equation}
\begin{split}
LMC~=~&A~+~B~\times~\frac{t}{\rm 1000days}~+~C~\times~(\frac{t}{\rm 1000days})^2\\
	&~+~D~\times~(\frac{t}{\rm 1000days})^3~+~E~\times~(\frac{t}{\rm 1000days})^4\\
	&~+~F~\times~\sin(\frac{2\pi t}{T_{QPOs}}~+~\phi_0)
\end{split}
\end{equation}
with $A~=~16.67\pm0.08$, $B~=~0.31\pm0.13$, $C~=~-0.11\pm0.07$, $D~=~0.02\pm0.01$, 
$E~=~-0.002\pm0.001$, $F~=~0.241\pm0.005$, $T_{QPOs}~=~1398\pm5$, $\phi_0~=~4.81\pm0.08$, 
leading to expected QPOs with a periodicity about $1398\pm5$ days (3.8 years). And 
the determined component described by the sinusoidal function and the component described by the 
four-degree polynomial function are also shown in top left panel of Figure~\ref{lmc}.

%%%3
	Before proceeding further, one point should be noted. The reason leading to large 
$\chi^2/dof~\sim9.3$ is mainly due to random fluctuations in the light curves. In order to show 
further evidence to support the sinusoidal component, only the polynomial function is applied to 
re-describe the light curves, leading to the calculated $\chi^{2}_{1}/dof_1~\sim4785.73/250\sim19.1$. 
Then, the F-test statistical technique is applied to determine whether the sinusoidal component 
is preferred. Based on the model functions with and without considerations of the sinusoidal 
component, the calculated $F_p$ value is about
\begin{equation}
	F_p=\frac{\frac{\chi^2_{1}-\chi^{2}}{dof_1-dof}}{\chi^2/dof}\sim88.8
\end{equation}
Based on $dof_1-dof$ and $dof$ as number of dofs of the F distribution numerator and denominator,
the expected value from the statistical F-test with confidence level quite higher than $8\sigma$ 
will be near to $F_p$. Therefore, the confidence level is higher than $8\sigma$ through the F-test 
statistical technique, to support the determined sinusoidal component is preferred.

%%%%4
	Meanwhile, based on the determined periodicity about $1398\pm5$ days, the phase folded 
light curve $LMC_{pf}$ shown in top right panel of Figure~\ref{lmc}, which can also be well 
described by a sinusoidal function with $\chi^2/dof~\sim~9.3$
\begin{equation}
LMC_{pf}~=~(-0.249\pm0.001)\times~\sin(2\pi t~-~(0.008\pm0.007))
\end{equation} 
Moreover, based on the F-test technique, the corresponding 99.9999\% confidence bands to the 
best fitting results are also shown in top panels of Figure~\ref{lmc}, and the residuals are 
shown bottom panels in Figure~\ref{lmc}, calculates by light curves minus the best fitting 
results. The directly best fitting results by the sinusoidal function to both the light curve 
and the phase folded light curve strongly support the optical QPOs in \obj.

\begin{figure*}
\centering\includegraphics[width = 8cm,height=5cm]{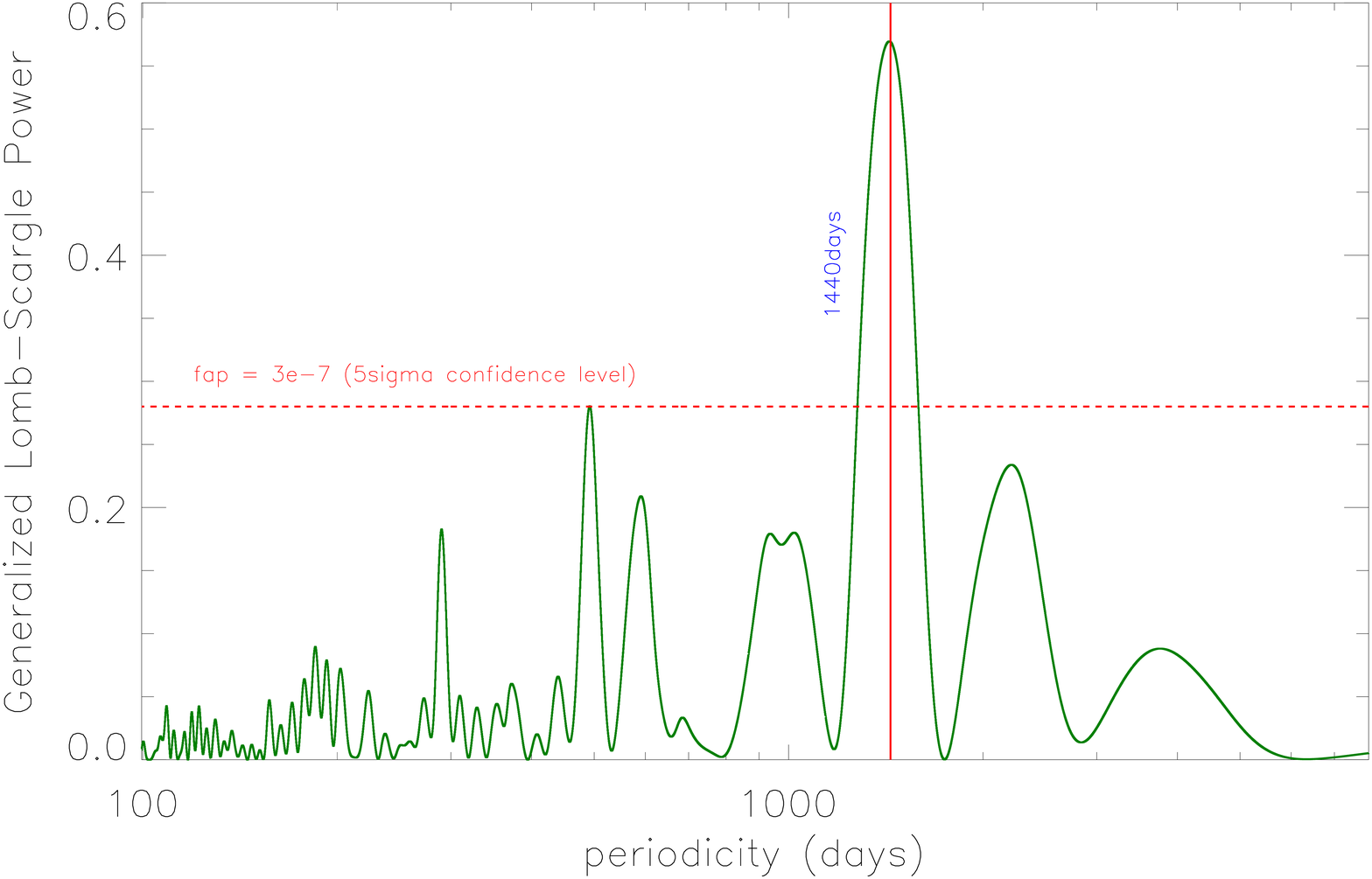}
\centering\includegraphics[width = 8cm,height=5cm]{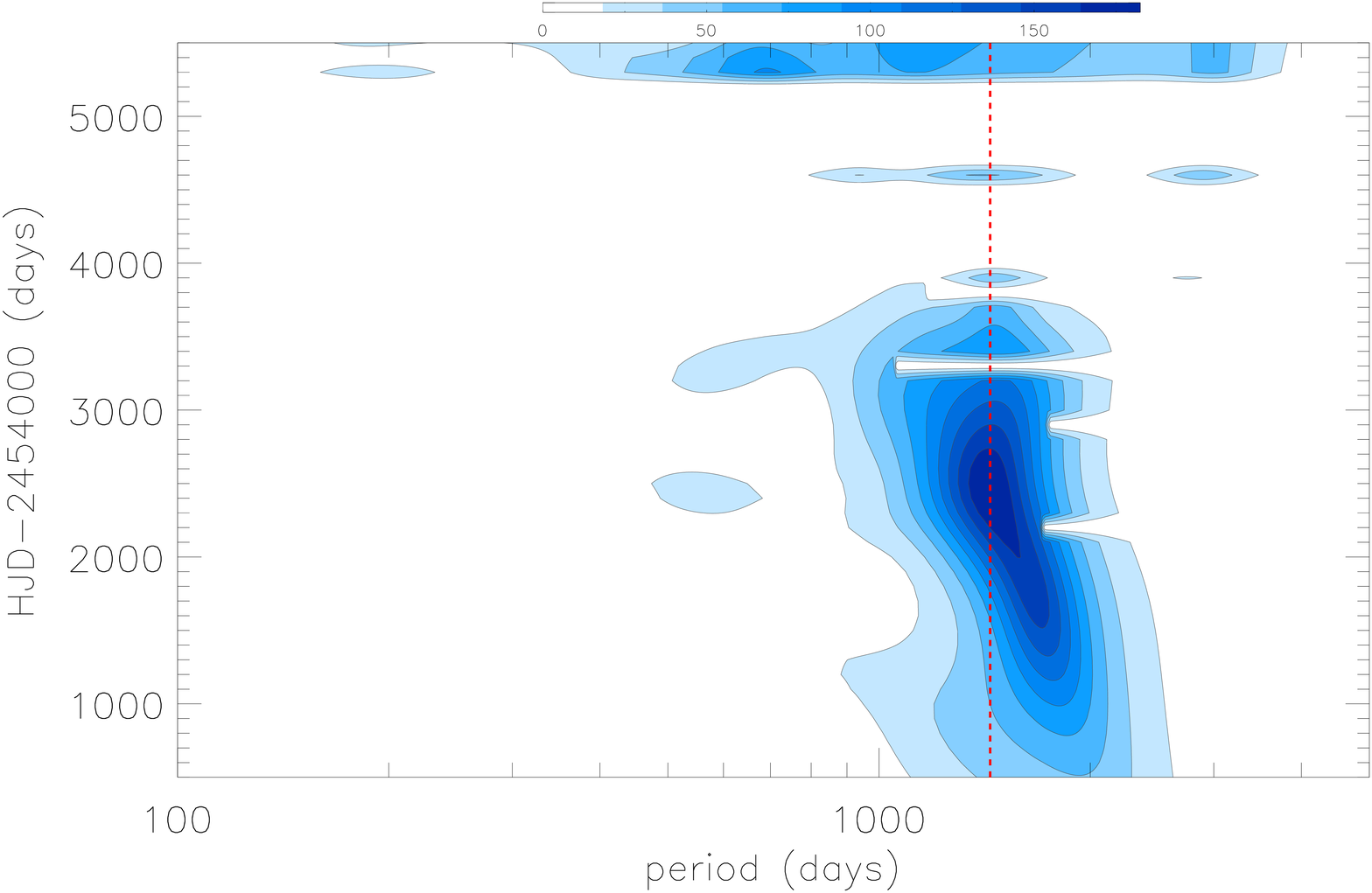}
\caption{Left panel shows results through the Generalized Lomb-Scargle periodogram. Horizontal 
dashed red line shows the 5$\sigma$ confidence level through the bootstrap method (the false-alarm 
probability of 3e-7). Vertical red line mark the peak around 1440days. Right panel shows the 
power maps in \obj~ determined by the WWWZ technique with frequency step of 0.00001 and searching 
periods from 100days to 5000days applied to the light curve (combining CSS V-band and ZTF $g$-band 
light curves) shown in left panel of Figure~\ref{lmc}. Vertical red dashed line marks the position 
with periodicity about 1440days. 
}
\label{qpo}
\end{figure*}

%%%5
	Besides the direct fitting results shown in Figure~\ref{lmc} by the sinusoidal function, 
the improved Generalized Lomb-Scargle (GLS) periodogram \citep{ln76, sj82, zk09, vj18} is applied 
to check the periodicities in the long-term combined CSS and ZTF variabilities in \obj, similar 
as what have been discussed in \citet{zb16} and as what we have done in \citet{zh22}. Although 
without considerations of the magnitude difference between CSS V-band and ZTF $gr$-bands, higher 
than 5$\sigma$ confidence level (the false-alarm probability of 3e-7) determined by the bootstrap 
method as discussed in \citep{ic19}, there is one periodicity $1440\pm16$days detected by the 
GLS periodogram shown in left panel of Figure~\ref{qpo}, showing quite similar 
periodicity as the determined 1400days in Figure~\ref{lmc}.

%%%%6
	In our more recent paper, \cite{zh22} have shown that the auto-correlation analysis 
(ACF) technique can be applied to determined QPOs, through created smoothly evenly-sampled light 
curves after interpolation technique applied to the observed light curve. However, there is a 
large time gap (also probable magnitude difference discussed in the following subsection) between 
the CSS V-band light curve and ZTF $gr$-band light curves, therefore QPOs in the light curve 
shown in Figure~\ref{lmc} is not appropriate to be determined by the ACF technique.

%%%7
	Then, the commonly accepted weighted wavelet z-transformation (WWZ) technique \citep{fg96, 
al16, gt18, ks20, ly21} can be applied to the combined light curve\footnote{The combined light 
curve from CSS V-band and ZTF $r$-band can lead to totally similar WWZ power maps.} from CSS V-band 
and ZTF $g$-band to determined QPOs in \obj, with the WWZ technique determined power maps shown in 
right panel of Figure.~\ref{qpo}. The WWZ technique determined periodicity is about 
$1400\pm110$days, with the uncertainty 110days determined by the bootstrap method as follows. 
Within 1000 different frequency steps randomly selected from 0.0000025 to 0.00025 applied in the 
WWZ technique, there are 1000 WWZ technique determined periodicities. Then, half width at half 
maximum of distribution of the 1000 WWZ technique determined periodicities is accepted as the 
uncertainty of the WWZ technique determined periodicity, showing quite similar periodicity as the 
determined values by the direct fitting results and by the GLS periodogram. Therefore, the periodicity 
about 1400days can be well accepted in \obj.

%%%8
	Finally, the QPOs with a periodicity around 1400days (3.8years) in \obj~ can be well 
detected from the 16.3yr-long combined photometric light curve (time duration about 4.3 times 
longer than the detected periodicity) with confidence level higher than 5$\sigma$, based on the 
best-fitting results directly by the sinusoidal function shown in the left panels of 
Figure~\ref{lmc}, on the sine-like phase-folded light curve shown in the right panels of 
Figure~\ref{lmc}, on the results of GLS periodogram shown in the left panel of Figure~\ref{qpo}, 
and on the power maps determined by the WWZ technique shown in right panel of Figure~\ref{qpo}.

\begin{figure*}
\centering\includegraphics[width = 18cm,height=6cm]{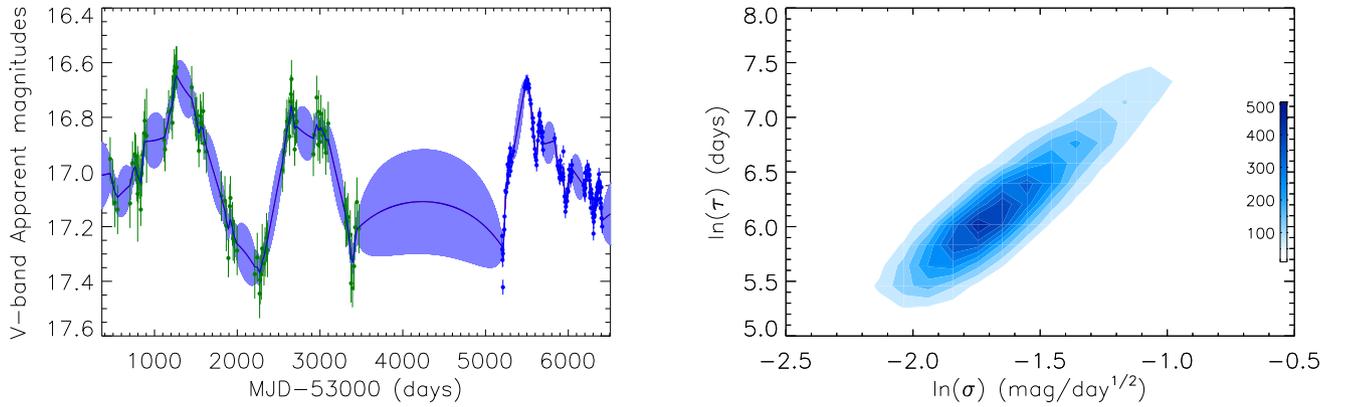}
\caption{Left panel shows the DRW-determined best descriptions to the long-term variabilities of 
\obj. Solid blue line and area filled with light blue show the best descriptions and the 
corresponding 1$\sigma$ confidence bands. Symbols in dark green and in blue represent the CSS V-band 
data points and the ZTF $g-$band data points. Right panel shows the MCMC determined two-dimensional 
posterior distributions of the DRW process parameters of $\ln(\tau)$ and $\ln(\sigma)$.
}
\label{drw}
\end{figure*}

\begin{figure*}
\centering\includegraphics[width = 18cm,height=6cm]{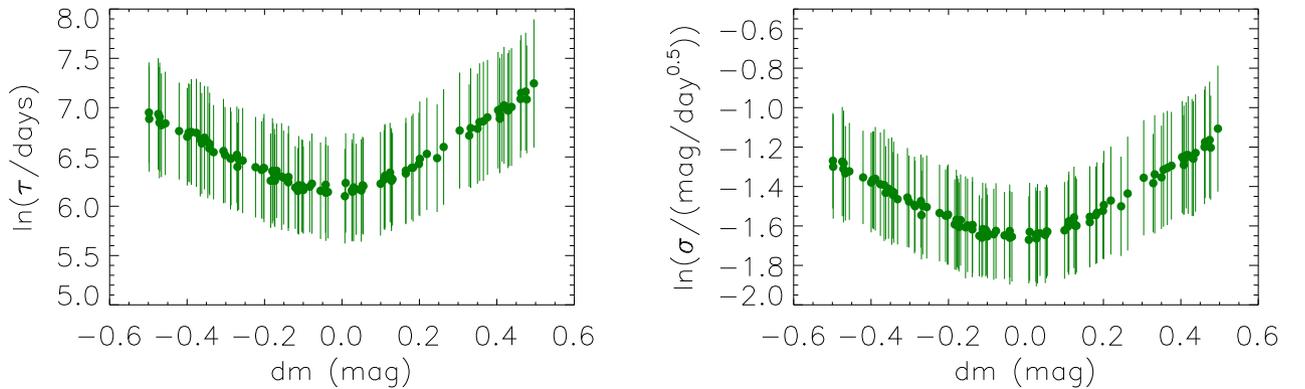}
\caption{Dependence of DRW process parameters of $\ln(\tau/days)$ (left panel) and 
$\ln(\sigma/(mag/day^{0.5}))$ (right panel) on the magnitude difference $dm$ between CSS light 
curve and the ZTF light curve. In each panel, solid circles plus error bars in dark green 
represent the determined DRW process parameters and corresponding uncertainties through the 
JAVELIN code.
}
\label{dm}
\end{figure*}

%%%%should be corrected by F-test applied to determined 999999\% confidence bands
\begin{figure*}
\centering\includegraphics[width = 18cm,height=10cm]{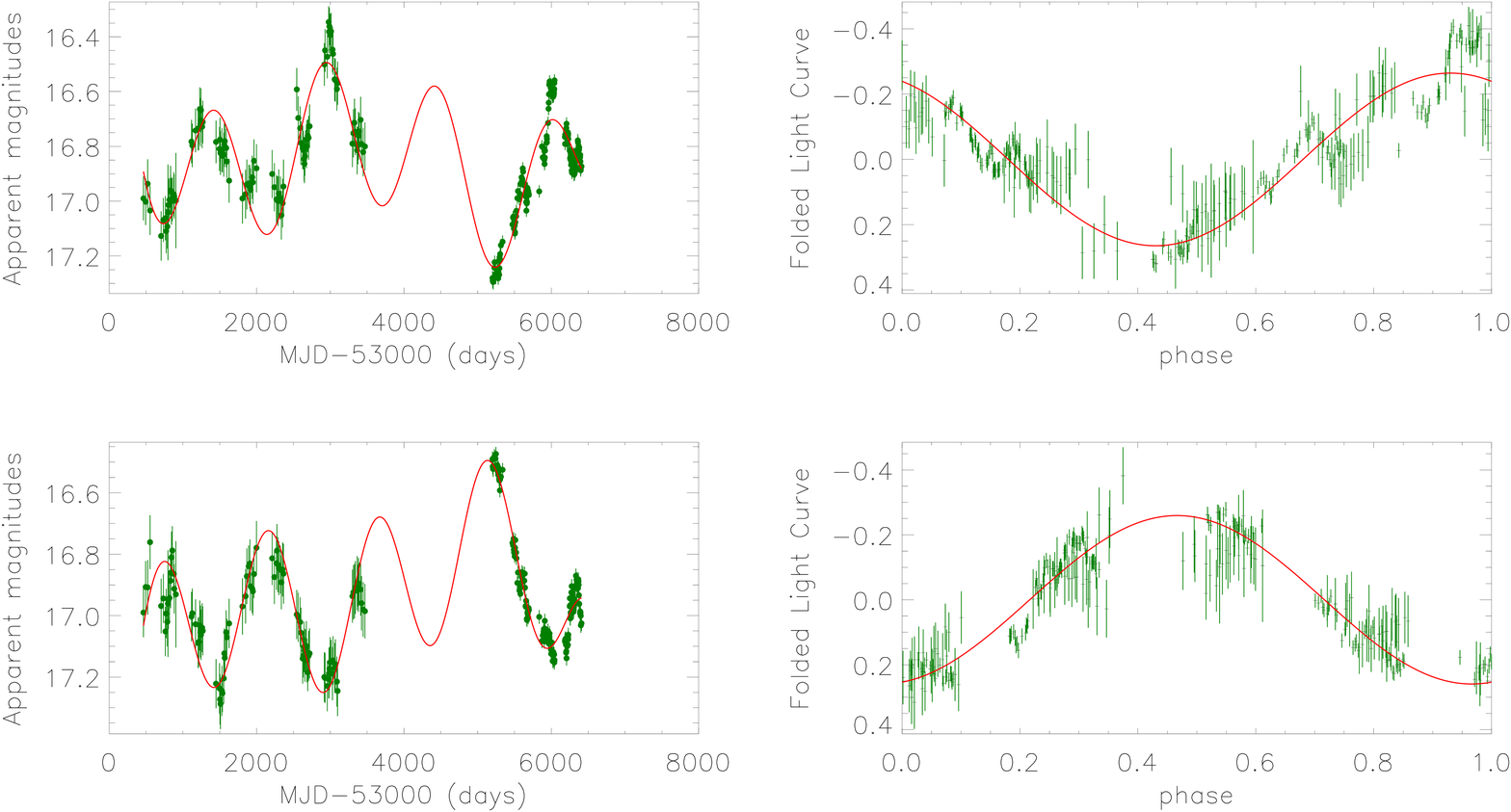}
\caption{Two examples on probable mis-detected QPOs in the simulating light curves by the CAR 
process. In each left panel, solid dark green circles plus error bars show the simulating light 
curve, solid red line shows the best descriptions to the light curve, based on a sinusoidal 
function plus a four-degree polynomial function. In each right panel, the corresponding phase 
folded light curve (dark green) and corresponding best fitting results (Solid red line) are shown.
}
\label{fake}
\end{figure*}

	Moreover, besides light curves from CSS and ZTF, long-term variabilities of \obj~ are 
also collected from Pan-STARRS (the Panoramic Survey Telescope And Rapid Response System) 
\citep{fm20, mc20}. There are large time gaps in Pan-STARRS $grizy$-band light curves, therefore, 
the Pan-STARRS light curves are not considered in Figure~\ref{lmc} and not to be described by 
the sinusoidal function. Here, solid purple triangles in top left panel of Figure~\ref{lmc} show 
the Pan-STARRS $z$-band light curve\footnote{Actually, the Pan-STARRS $griy$-band light curves 
can also follow the best fitting results shown in top left panel of Figure~\ref{lmc}.} with 17 
data points (more data points than the other Pan-STARRS band light curves), which are well 
consistent with the sinusoidal function described best fitting results to the CSS and ZTF light 
curves. Therefore, the Pan-STARRS light curves can be applied as additional evidence to support 
the optical QPOs in \obj, but there are no further discussions on variabilities of Pan-STARRS 
light curves.

\subsection{Mis-detected QPOs related to central intrinsic AGN activities?}

	In the subsection, it is necessary and interesting to determine whether the determined 
optical QPOs was mis-detected QPOs tightly related to central intrinsic AGN activities of \obj, 
although the different techniques applied in the subsection above can lead to the detected optical 
QPOs in \obj. Similar as what we have done in \citet{zh22} to check probability of mis-detected 
QPOs in SDSS J075217.84+193542.2, the following procedure is applied to determine whether are the 
detected QPOs in \obj~ mis-detected QPOs in light curves related to central AGN activities well 
described by damped random walk process.

%%%2
	Variability is fundamental characteristics of AGN \citep{mr84, um97, ms16, bg20, bs21} 
which have been proved to well described by the well-applied Continuous AutoRegressive process 
(CAR process or the improved damped random walk process (DRW process)) \citep{kbs09, koz10, zk13, 
kb14, sh16, zk16, zh17, tm18, mv19, sr22}. Here, the DRW process is also applied to describe the 
long-term variabilities of \obj, through the public code JAVELIN (Just Another Vehicle for 
Estimating Lags In Nuclei) \citep{koz10, zk13} with two process parameters of intrinsic 
characteristic variability amplitude and timescale of $\sigma$ and $\tau$ applied in the exponential 
covariance matrix $S$ of time-dependent variabilities $S_{ij}=\frac{1}{2}\tau\sigma^2exp(-|t_i-t_j|/\tau)$. 
Commonly, the parameter $\tau$ called as “relaxation time” is tightly related to timescales of 
central AGN accreting process, and the parameter $\sigma$ represents variability resulting from 
local random deviations in central accretion disk structures. The best descriptions to the light 
curve are shown in the left panel of Figure~\ref{drw}. And the corresponding MCMC (Markov Chain 
Monte Carlo) \citep{fh13} determined two dimensional posterior distributions of the parameters 
of $\sigma$ and $\tau$ are shown in right panel of Figure~\ref{drw}, with the determined 
$\ln(\tau/days)\sim6.22\pm0.59$ ($\tau\sim501$days) and $\ln(\sigma/(mag/dyas^{1/2}))\sim-1.63\pm0.21$ 
($\sigma\sim0.19{\rm mag/day^{1/2}}$). Comparing with the long-term variabilities of SDSS quasars 
shown in Figure~3\ in \citet{mi10}, the DRW determined 
$\log(SF_{\infty}/mag)=\log(\sigma\times\sqrt{\tau})\sim0.64$ is definitely one magnitude larger 
than the mean value around -0.7 of the SDSS quasars, indicating \obj~ is an interesting target.

%%%3
	Certainly, there are no considerations of magnitude difference between CSS light curve 
$[t_{css},~L_{CSS}]$ and ZTF light curve $[t_{ZTF},~L_{ZTF}]$ on the results above. Here, 
considering randomly selected magnitude difference $dm$ from -0.5 to 0.5, new light curves 
$[t_{new},~L_{new}]$ can be well determined as 
\begin{equation}
\begin{split}
	t_{new}~&=~[t_{css},~t_{ZTF}]  \\
	L_{new}~&=~[L_{CSS},~L_{ZTF}+dm]
\end{split}
\end{equation}
in order to check effects of probable magnitude difference on DRW process parameters in \obj. 
Then, the same JAVELIN code is applied to describe the simulating 100 light curves of 
$[t_{new},~L_{new}]$ with 100 randomly selected $dm$. Dependence of the 100 determined $\sigma$ 
and $\tau$ on the parameter $dm$ are shown in Figure~\ref{dm}. It is clear that the magnitude 
difference between CSS light curve and ZTF light curve has apparent effects on the estimated 
DRW process parameters of $\sigma$ and $\tau$. However, considering the results in 
Figure~\ref{dm}, the determined $\log(SF_{\infty}/mag)\sim0.64$ with $dm~\sim~0$ is the 
smallest value among the simulating light curves with $dm~\ne~0$, to re-support that \obj~ is 
an interesting target. Here, quite larger $dm$ can lead to more apparently larger 
$\log(SF_{\infty}/mag)$. Therefore, no further discussions are given on larger values of 
$dm$ in the manuscript.

%%%4
	Then, probability of mis-detected QPOs from DRW process described intrinsic AGN 
variabilities can be estimated as follows. Based on the CAR process discussed in \citet{kbs09}:
\begin{equation}
\dif LMC_t=\frac{-1}{\tau}LMC_t\dif t+\sigma_*\sqrt{\dif t}\epsilon(t)~+~16.99
\end{equation}
where $\epsilon(t)$ is a white noise process with zero mean and variance equal to 1. Here, 
the mean value of $LMC_t$ is set to be 16.99 (the mean value of 
the light curve of \obj), which has no effects on the following results. Then, a series of 
10000 simulating light curves [$t$,~$LMC_t$] are created, with $\tau$ randomly selected from 
500days to 1500 days (the $\tau$ range shown in left panel of Figure~\ref{dm} of the 100 
light curves after considerations of $dm$) and $\tau\sigma_*^2/2$ randomly selected from 
0.03 to 0.096 (the variance range of the 100 light curves after considerations of $dm$) 
(the parameter $\sigma_*$ in unit of mag in the CAR process in \citet{kbs09} slightly 
different from the JAVELIN determined $\sigma$). And, time information $t$ are the same as 
the observational time information shown in left panel of Figure~\ref{lmc}. And the similar 
uncertainties $LMC_{t,~err}$ are simply added to the simulating light curves $LMC_t$ by
\begin{equation}
LMC_{t,~err}~=~LMC_{t}\times\frac{L_{err}}{L_{obs}}
\end{equation}
with $L_{obs}$ and $L_{err}$ as the observational CSS and ZTF light curves and the 
corresponding uncertainties shown in top left panel of Figure~\ref{lmc}.

%%%5
	Then, the following three simple criteria are applied to check whether QPOs can 
be detected in the simulating light curves. First, GLS-determined periodicities should 
be around 1400days (larger than 1250days and smaller than 1550days) with significance 
level higher than 5$\sigma$. Second, the simulating light curves can be best described 
by the equation (1) with $\chi^2/dof<15$ ($\chi^2/dof\sim9.3$ for the results shown in 
left panel of Figure~\ref{lmc}), and the best-fitting procedure determined periodicity 
is around 1400days (larger than 1250days and smaller than 1550days). Third, the 
corresponding phase folded light curve with subtractions of polynomial component can be 
well described by a sinusoidal function with $\chi^2/dof<15$ ($\chi^2/dof\sim9.3$ for the 
results shown in right panel of Figure~\ref{lmc}). Here, considering uncertainty 
of 5days for the determined periodicity through the best fitting results by sinusoidal 
function, uncertainty of 16days for the GLS periodogram determined periodicity, and 
uncertainty of 110days for the WWZ technique determined periodicity, the uncertainty about 
5days+16days+110days (about 131days) is accepted as the uncertainty of the determined 
periodicity 1400days. Then, the uncertainty of 131days leads to the accepted 
periodicity range from 1250days (a bit smaller than 1400-131days) to 1550days (a bit 
larger than 1400+131days) in the first criterion and in the second criterion. Finally, 
among the 10000 simulating light curves, there are 7 light curves with expected mis-detected 
QPOs with periodicity around 1400days, accepted the three criteria above. Moreover, 
Figure~\ref{fake} shows 2 of the 7 simulating light curves with mis-detected QPOs and 
the corresponding best-fitting results by equation (1) and the corresponding results on 
phase folded light curves. The results indicate that the DRW process (or the CAR process) 
can lead to light curves with mathematical determined QPOs (the mis-detected QPOs, or the 
fake QPOs), however, the probability of the mis-detected QPOs in CAR-process simulating 
light curves is around 0.07\% (7/10000). The results strongly indicate that the probability 
higher than 99.93\% (1-0.07\%) to support that the detected optical QPOs in \obj~ are not 
mis-detected QPOs from a pure CAR process described light curve.

	Furthermore, one additional point is noted. When the simulating $LMC_t$ are created 
above, the parameters of $\tau$ and $\sigma_*$ are randomly selected. If the parameters of 
$\tau$ and $\sigma_*$ are fixed, are there different results? Then, a new series $LMC_t$ 
are created as follows. The DRW process parameter $\tau$ is fix to be 501days, the value 
determined through the light curve without considerations of magnitude difference $dm$ 
between CSS light curve and ZTF light curve, and the value $\sigma_*\sim0.011$ is determined 
by the variance $\tau\sigma_*^2/2$ of CAR process created light curves to be 0.031 (the 
variance of the light curve of \obj~ shown in left panel of Figure~\ref{lmc}). Then, another 
10000 light curves are created by the CAR process. And then, based on the same three criteria, 
among the 10000 simulating light curves, there are 9 light curves with expected mis-detected 
QPOs. The results strongly indicate that different input parameters of $\tau$ and $\sigma_*$ 
have tiny effects on the final results, and that the probability higher than 99.9\% (smaller 
than 1-0.07\%, also smaller than 1-0.09\%) (significance level higher than 
3$\sigma$) to support that the detected optical QPOs in \obj~ are not mis-detected QPOs 
from a pure CAR process described variabilities related to central AGN activities.

\begin{figure}
\centering\includegraphics[width = 8cm,height=5cm]{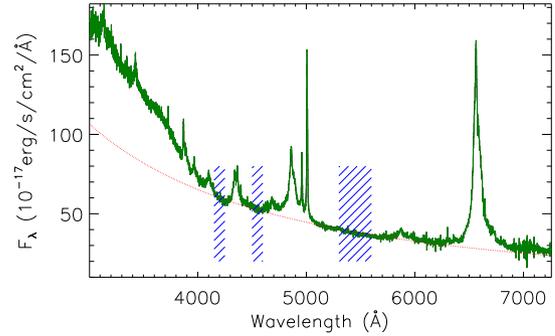}
\caption{The galactic reddening corrected spectrum of \obj~ in rest frame. The dotted red 
line represents the determined power-law continuum emissions. The areas filled with green 
lines show the wavelength windows applied to determine the power law continuum emissions.
}
\label{spec}
\end{figure}

\begin{figure*}
\centering\includegraphics[width = 8cm,height=6cm]{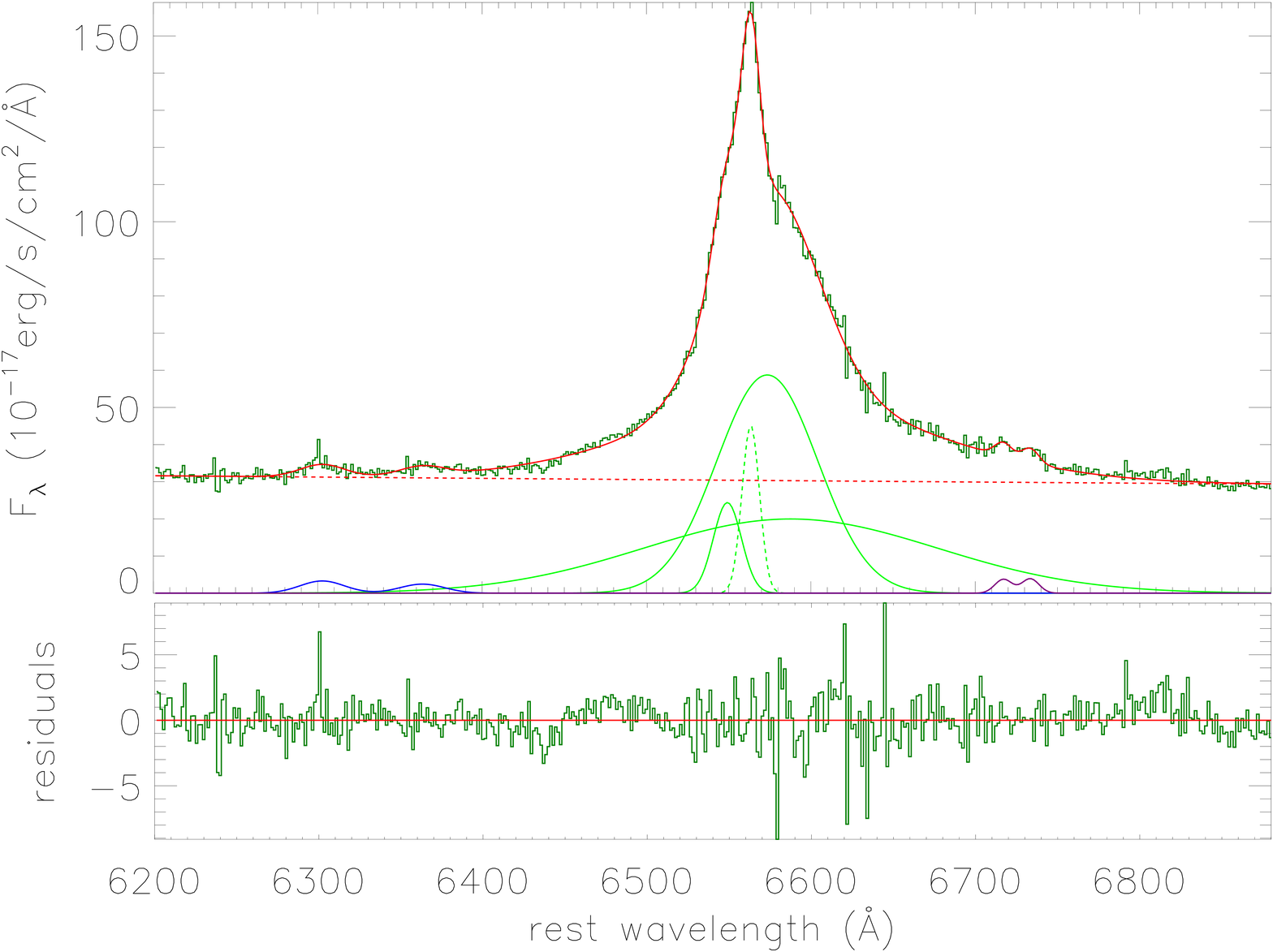}
\centering\includegraphics[width = 8cm,height=6cm]{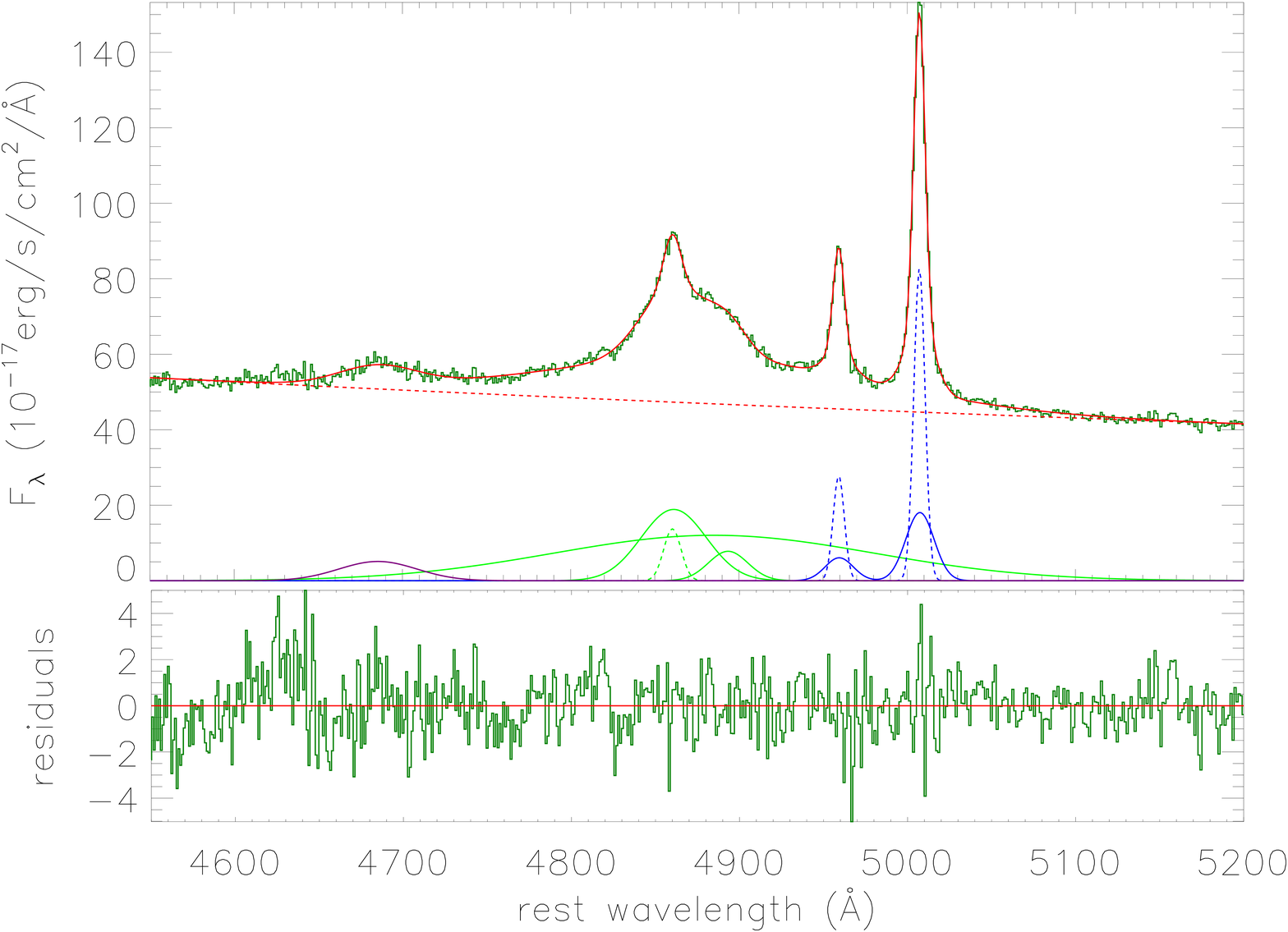}
\caption{Left panels show the best-fitting results (top panel) and the corresponding residuals 
(bottom panel) to the emission lines around the H$\alpha$. Right panels show the best-fitting 
results (top panel) and the corresponding residuals (bottom panel) to the emission lines around 
the H$\beta$. In each top panel, solid dark green line shows the SDSS spectrum, solid red line 
shows the best fitting results, solid green lines show the determined three broad Gaussian 
components in broad Balmer line, dashed green line shows the determined narrow Gaussian component 
in narrow Balmer line, dashed red line shows the determined power law continuum emissions. 
In top left panel, solid blue lines show the determined [O~{\sc i}] doublet, solid purple 
lines show the determined [S~{\sc ii}] doublet. In top right panel, dashed blue and solid blue 
lines show the determine core and extended components in the [O~{\sc iii}] doublet, solid purple 
line shows the determined broad He~{\sc ii} line. In each bottom panel, solid red line shows 
residuals=0.
}
\label{ha}
\end{figure*}

\begin{figure*}
\centering\includegraphics[width = 8cm,height=6cm]{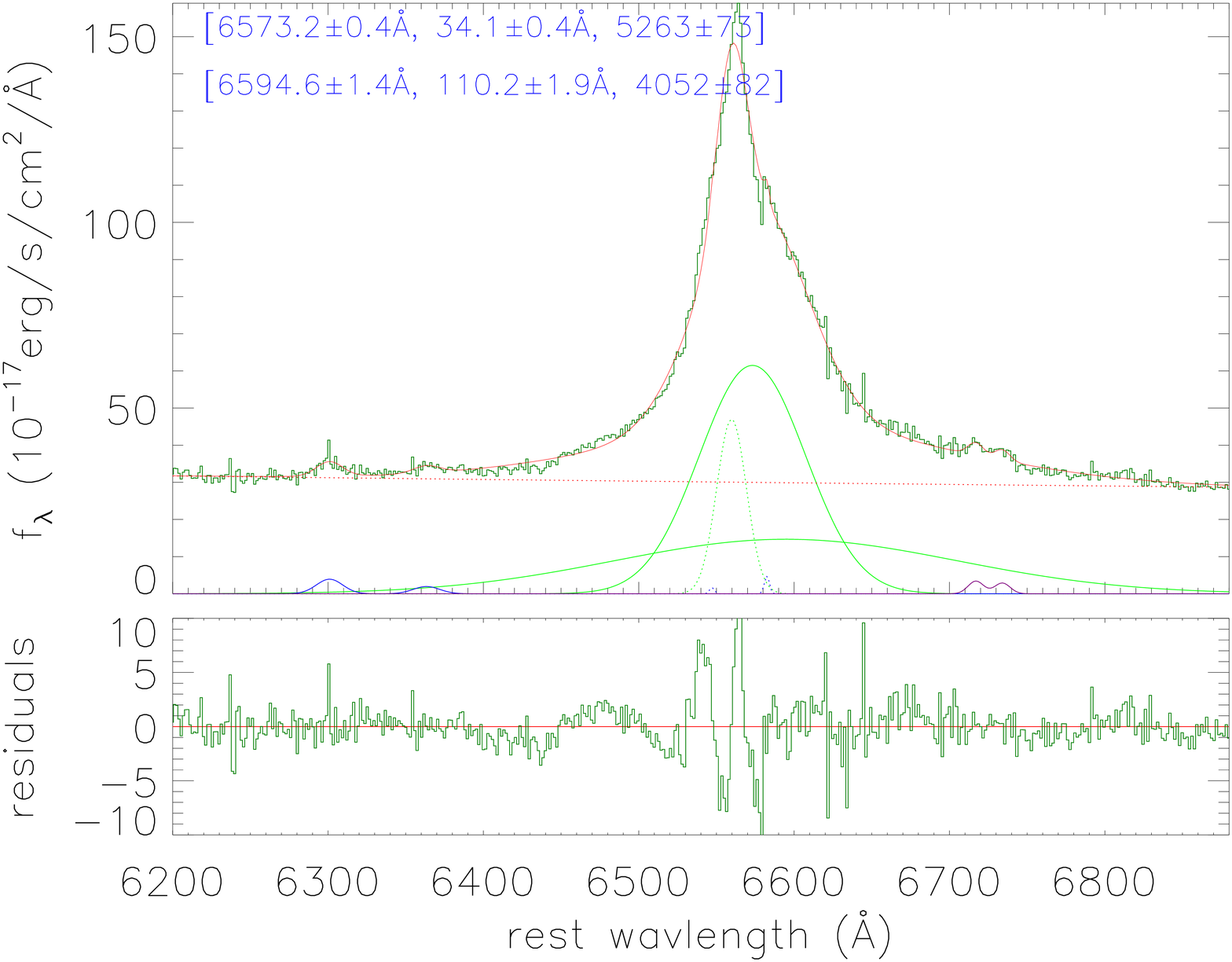}
\centering\includegraphics[width = 8cm,height=6cm]{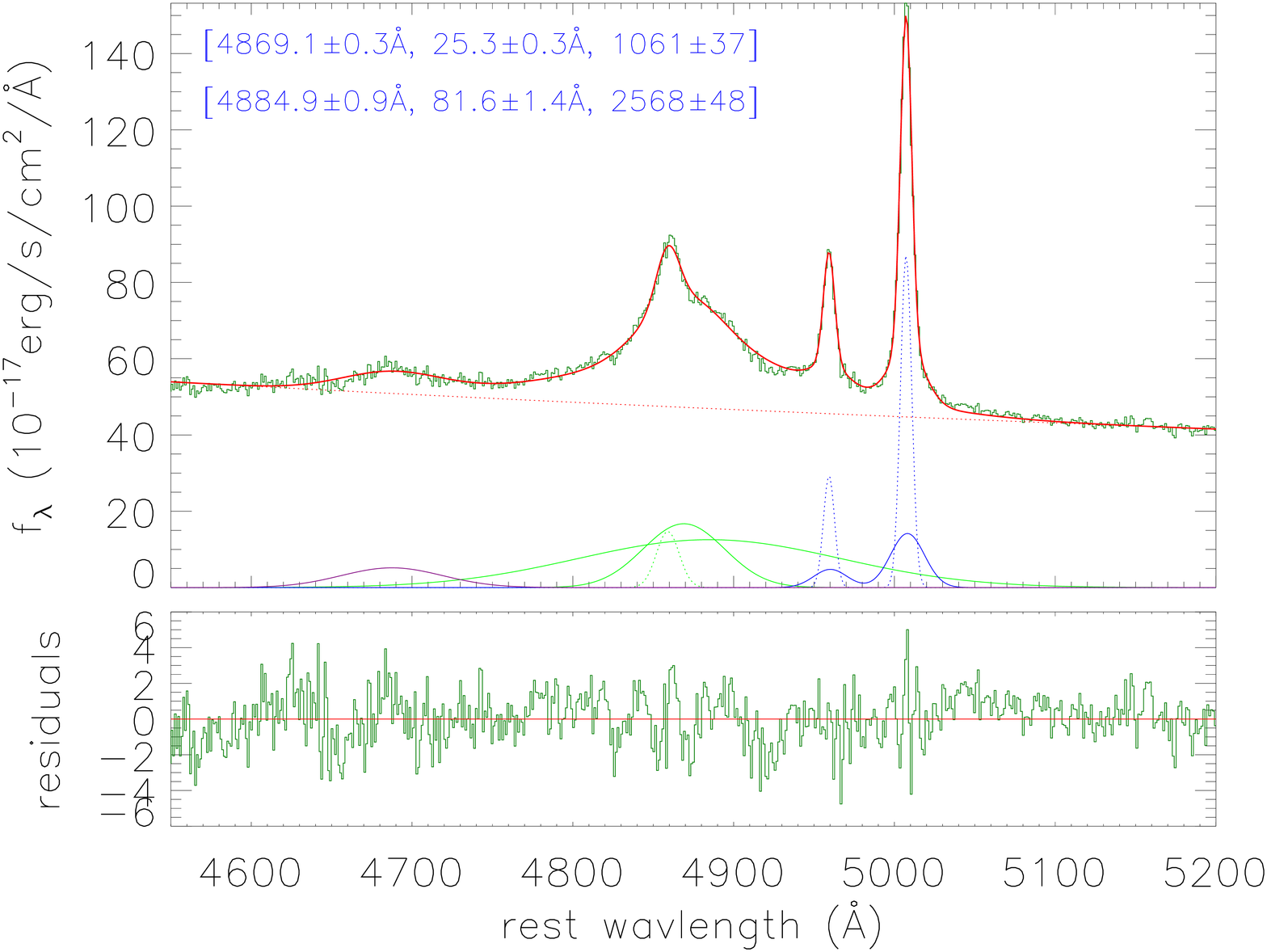}
\caption{Similar as the results shown in Figure~\ref{ha}, but for each broad Balmer component 
described by two broad Gaussian functions. The determined parameters of each of the two broad Gaussian 
components are marked in blue characters with [central wavelength in unit of \AA, second moment in unit 
of \AA, flux in unit of $10^{-17}{\rm erg/s/cm^2/\AA}$].}
\label{hab}
\end{figure*}

\begin{table}
\caption{Line parameters}
\begin{tabular}{llll}
\hline\hline
line & $\lambda_0$ & $\sigma$ & flux \\
\hline\hline
\multirow{3}{*}{Broad H$\alpha$}  &  6587.8$\pm$1.3  & 88.9$\pm$2.3  & 4478$\pm$93 \\
	& 6573.4$\pm$0.4 & 30.7$\pm$0.5 & 4525$\pm$112 \\
	& 6549.2$\pm$1.9 & 8.6$\pm$1.3 & 523$\pm$110 \\
\hline
\multirow{3}{*}{Broad H$\beta$}  &  4885.1$\pm$1.7  & 95.3$\pm$2.1  & 2874$\pm$59 \\
        & 4861.2$\pm$2.6 & 19.4$\pm$1.5 & 916$\pm$123 \\
	& 4893.6$\pm$2.9 & 11.7$\pm$2.1 & 229$\pm$101 \\
\hline
Broad He~{\sc ii} & 4685.4$\pm$1.4 & 23.2$\pm$1.5 & 297$\pm$20 \\
\hline
Narrow H$\alpha$ & 6563.4$\pm$0.5 & 5.3$\pm$0.4 &  602$\pm$100 \\
\hline
Narrow H$\beta$ & 4860.4$\pm$0.4 & 5.2$\pm$0.6 &  178$\pm$34 \\
\hline
\multirow{2}{*}{[O~{\sc iii}]$\lambda5007$\AA} & 5007.3$\pm$0.1 & 3.4$\pm$0.1 & 696$\pm$36 \\
& 5007.5$\pm$0.3 & 8.4$\pm$0.6 & 382$\pm$33 \\
\hline
[O~{\sc i}]$\lambda6300$\AA & 6302.7$\pm$1.5 & 13.4$\pm$1.2 &  111$\pm$13 \\
\hline
[O~{\sc i}]$\lambda6363$\AA & 6363.7$\pm$1.9 & 13.6$\pm$1.3  & 83$\pm$11 \\
\hline
[S~{\sc ii}]$\lambda6716$\AA & 6717.2$\pm$1.1 & 5.2$\pm$1.1 &  49$\pm$8 \\
\hline
[S~{\sc ii}]$\lambda6731$\AA & 6733.5$\pm$1.1 & 5.3$\pm$1.2 & 51$\pm$8 \\
\hline\hline
\end{tabular}\\
Notice: The first column shows which line is measured. The Second, third, fourth columns show the 
measured line parameters: the center wavelength $\lambda_0$ in unit of \AA, the line width (second 
moment) $\sigma$ in unit of \AA~ and the line flux in unit of ${\rm 10^{-17}~erg/s/cm^2}$. \\ 
For the broad H$\alpha$ (broad H$\beta$), there are three Gaussian components.  \\
For [O~{\sc iii}]$\lambda5007$\AA, there are two components: one core component and one extended 
component.
\end{table}

\section{Spectroscopic properties of \obj}

%1
	Figure~\ref{spec} shows the high-quality galactic reddening corrected ($A_B=0.102$) 
spectrum of \obj~ with PLATE-MJD-FIBERID=0526-52312-0537 collected from SDSS DR16 \citep{ap21}. 
The apparently blue continuum emissions lead \obj~ to be well classified as a SDSS quasar. The 
featureless continuum emissions can be well described by a power law function, 
$f_\lambda~\propto~\lambda^{-1.70}$ through the following three windows [4150\AA,~4250\AA], 
[4500\AA,~4600\AA], [5300\AA,~5600\AA], with continuum luminosity at rest wavelength 5100\AA~ 
to be about $\lambda~L_{\rm 5100}~=~4.94~\times~10^{45}~{\rm erg/s}$.

%%%2
	Considering the optical QPOs tightly related to an expected central BBH system in \obj, 
it is interesting to check broad emission line properties. Emission lines around H$\alpha$ in 
\obj~ can be well measured within rest wavelength range from 6200\AA~ to 6850\AA. Slightly 
different from what we have recently done in \citet{zh21b, zh21c}, three broad and one narrow 
Gaussian functions are applied to describe the broad and narrow H$\alpha$ components, due to 
complicated Balmer emission features. And six another Gaussian functions are applied to describe 
[O~{\sc i}], [N~{\sc ii}] and [S~{\sc ii}] doublets. And, a power law function is applied to 
describe the AGN continuum emissions. Left panels of Figure~\ref{ha} shows the best fitting 
results with $\chi^2/dof\sim1.27$ (summed squared residuals divided by degree of freedom) 
to the emission lines around H$\alpha$ and the corresponding residuals (spectrum minus the best 
fitting results), through the Levenberg-Marquardt least-squares minimization technique. When the 
fitting procedure is running, only one restriction is accepted that the emission flux of each 
Gaussian emission component is not smaller than zero. Actually, as the shown best-fitting results 
in top left panel of Figure~\ref{ha}, there are no [N~{\sc ii}] emission features, due to 
measured line intensities near to zero. The measured line parameters are listed in Table~1.

%%%4
	Meanwhile, right panels of Figure~\ref{ha} shows the best fitting results (with 
$\chi^2/dof~\sim~1.13$) to the emission lines around H$\beta$ with rest wavelength from 4450\AA~ 
to 5350\AA~ by the following model functions through the same Levenberg-Marquardt least-squares 
minimization technique, to support the similar line profiles between broad H$\alpha$ and broad 
H$\beta$. There are three broad and one narrow Gaussian components applied to describe the broad 
and narrow H$\beta$, four Gaussian components applied to describe the core and extended components 
in [O~{\sc iii}] doublet, one Gaussian function applied to describe the He~{\sc i} line, and 
a power law component applied to describe the continuum emissions underneath the emission lines. 
When the fitting procedure is running, the flux ratios of the components of the [O~{\sc iii}] 
doublet are set to the theoretical values of 3. The measured line parameters of the emission 
lines around H$\beta$ are also listed in Table 1.

	Before proceeding further, model functions considering two broad Gaussian components in 
broad Balmer line are also applied to describe emission lines around H$\alpha$ and around H$\beta$. 
The determined fitting results are shown in Figure~\ref{hab} with $\chi^2/dof\sim1.31$. The residuals 
especially around H$\alpha$ shown in bottom left panel of Figure~\ref{hab} indicate that the model 
functions considering two broad Gaussian components in broad H$\alpha$ are not appropriate to 
describe the emission lines around H$\alpha$. Furthermore, based on the model functions with two 
or three broad Gaussian functions to describe broad Balmer line, similar as what have been done above 
in the subsection 2.1, the calculated $F_p$ value about $F_p\sim29.5$ can be applied in the F-test 
statistical technique to confirm model functions considering three broad Gaussian components in 
broad Balmer line are preferred with confidence level quite higher than $8\sigma$ (3 and 571 as number 
of dofs of the F distribution numerator and denominator). Moreover, based on the fitting results 
shown in Figure~\ref{hab} considering two broad Gaussian components in broad Balmer line, both the 
two determined broad Gaussian components are red-shifted components with central 
wavelengths about $6573.2\pm0.4$\AA~ and $6594.6\pm1.4$\AA~ ($4869.1\pm0.3$\AA~ and $4884.9\pm0.9$\AA) 
in broad H$\alpha$ (in broad H$\beta$), not leading to a blue-shifted broad component plus a 
red-shifted broad component. Therefore, emission line parameters are mainly considered from the 
best fitting results by the model functions considering three broad Gaussian components in broad 
Balmer line in the manuscript.

	It is clear that there are complicated line profiles of the broad Balmer emission lines. 
It is hard to found two apparent broad components in broad Balmer lines, probably indicating the 
central two systems, each system including its own BH accreting structures and related BLRs, are 
nearer enough. Therefore, it is hard to test BBH model expected emission line features.

	Based on the determined multiple Gaussian broad components included in line profiles 
$f_\lambda$ of Balmer lines, basic parameters of central wavelength $\lambda_0$, line width 
(second moment $\sigma_0$) and line luminosity $L$ of broad Balmer lines can be estimated by 
\begin{equation}
\begin{split}
&\lambda_0~=~\frac{\int\lambda f_\lambda\dif\lambda}{\int~f_\lambda\dif\lambda} \ \ \ \ \ \ \ 
\ \ \ 	\sigma_0^2~=~\frac{\int\lambda^2f_\lambda\dif\lambda}{\int~f_\lambda\dif\lambda}~-~\lambda_0^2 \\ 
&L~=~4\pi~D^2~(\int~f_\lambda\dif\lambda)
\end{split}
\end{equation}
with $D$ as the distance between the Earth and \obj, leading ($\lambda_0$, $\sigma_0$, $L$) to 
be (6577.8$\pm$1.9\AA, 63.7$\pm$2.3\AA, $(2.29\pm0.11)\times10^{43}{\rm erg/s}$) and 
(4879.3$\pm$2.9\AA, 80.2$\pm$2.1\AA, $(9.33\pm0.16)\times10^{42}{\rm erg/s}$) of broad H$\alpha$ 
and broad H$\beta$, respectively. The determined line parameters can be applied to determine 
virial BH mass in \obj~ in the following section.

	Moreover, based on the determined Balmer line luminosity, an interesting result can be 
found. As discussed in \citet{gh05}, there is a strong linear correlation between continuum 
luminosity and Balmer line luminosity in SDSS quasars. However, in \obj, based on the continuum 
luminosity at 5100\AA~ about $4.94\times10^{45}{\rm erg/s}$, the expected H$\alpha$ line 
luminosity is about $4.79\times10^{44}{\rm erg/s}$, about 20 times larger than the determined 
H$\alpha$ luminosity (even considering the narrow Gaussian component in narrow H$\alpha$) in 
\obj, which is an interesting point. Combining the blue quasar-shape continuum emissions, 
and after checking flux ratio around 2.4 of total H$\alpha$ to total H$\beta$, there are few 
effects of dust obscurations on the smaller observed H$\alpha$ luminosity, probably indicating 
special central physical properties in \obj~ which will be tested by broad line variabilities 
in the future.

%The smaller observed H$\alpha$ luminosity than the intrinsic 
%expected H$\alpha$ luminosity actually can not explained by simple dust obscurations on whole BH 
%accreting system plus the broad emission line regions, because dust obscurations have more 
%serious effects on continuum luminosity at 5100\AA~ than on H$\alpha$ luminosity. At current 
%stage, an oversimplified viewpoint can be proposed that dust clouds have effects only on both 
%or one of the two broad emission line regions in the BBH system, therefore, leading to weaker 
%H$\alpha$ luminosity but dew effects on continuum luminosities. More detailed evidence to support 
%
%80.160162       4879.3336       4196.4504
%       81.806600       4880.1688       4018.6394

\section{Main Discussions}

%%%1
%	The determined two broad H$\alpha$ components provide better chances to check whether is 
%there a central BBH system in SDSS J0752, because the two BH masses in the probable BBH system can 
%be well determined by properties of the two broad components. 

%%%1
	Not similar as well determined blue-shifted and red-shifted broad Balmer components 
in the BBH candidate in SDSS J075217.84+193542.2 reported in \citet{zh22}, there are no well determined two 
broad components in broad Balmer lines in \obj. It is hard to estimate the two BH masses of 
central BBH system, through properties of two separated broad components in Balmer lines applied 
in Virial equations \citep{pe04, vp06, sh11}. However, under the assumption of BBH system in \obj, 
upper limit of central total BH mass and corresponding upper limit of space separation between 
the central two BHs can be simply estimated as follows.

%%%%2
	In order to ignore effects of complicated line profiles of broad Balmer lines, the correlation 
between BH mass and continuum luminosity reported in \citet{pe04} (see their Equation 9) is applied to 
determine the upper limit of central total BH mass without further assumptions, rather than virial 
equations with applications of line/continuum luminosity and broad line widths. Considering $\lambda~L_{c1}$ 
and $\lambda~L_{c2}$ as continuum luminosity at 5100\AA~ from each BH accreting system in the assumed 
BBH system in \obj, each BH mass can be estimated as
\begin{equation}
\begin{split}
M_{\rm BH,~1}~&=~7.58\times10^7{\rm M_\odot}(\frac{\lambda~L_{c1}}{10^{44}{\rm erg/s}})^{0.79} \\
M_{\rm BH,~2}~&=~7.58\times10^7{\rm M_\odot}(\frac{\lambda~L_{c2}}{10^{44}{\rm erg/s}})^{0.79}
\end{split}
\end{equation}
Considering the observed continuum luminosity at 5100\AA~ about 
$\lambda~L_{\rm 5100}~=~4.94~\times~10^{45}~{\rm erg/s}$ larger than $\lambda~L_{c1}$ and also than 
$\lambda~L_{c2}$, upper limit of total BH mass ($M_{\rm BH}~=~M_{\rm BH,~1}+M_{\rm BH,~2}$) should be
\begin{equation}
\begin{split}
M_{\rm BH}~&=~M_{\rm BH,~1}+M_{\rm BH,~2}\\
	&~=~7.58\times10^7{\rm M_\odot}\times
	((\frac{\lambda~L_{c1}}{10^{44}{\rm erg/s}})^{0.79}~+~
	(\frac{\lambda~L_{c2}}{10^{44}{\rm erg/s}})^{0.79})\\
	&~<~7.58\times10^7{\rm M_\odot}~\times~2\times(\frac{\lambda~L_{5100}}{10^{44}{\rm erg/s}})^{0.79}\\
	&~\sim~3.3\times10^9{\rm M_\odot}
\end{split}
\end{equation}
Then, based on the upper limit of total BH mass in the central BBH system, upper limit of space separation between
the central two BHs can be estimated as
\begin{equation}
A_{BBH}=0.432\times M_{8}\times(\frac{P_{BBH}/year}{2652M_{8}})^{2/3}~\leq~0.018pc
\end{equation}
with $M_{8}$ as total BH mass of the BBH system in unit of $10^8{\rm M_\odot}$ and $P_{BBH}\sim3.8yr$ as
orbital period of the BBH system.

%%%3
	Meanwhile, besides the BBH system, precessions of emission regions with probable hot spots for 
the optical continuum emissions can also be applied to describe the detected optical QPOs in \obj. 
Considering line width (second moment) of broad Balmer lines and well measured continuum luminosity, 
virial BH mass in \obj~ can be estimated through the formula discussed in \citet{pe04} 
\begin{equation}
\frac{M_{\rm BH}}{\rm M_\odot}~=~5.5~\times~ \frac{\sigma^2_{\rm broad, H\alpha}
	~\times~ R_{\rm BLRs}}{G}\sim2.68\times10^9
\end{equation}
where $\sigma_{\rm broad, H\alpha}\sim2910~{\rm km/s}$ represents the second moment of the total broad 
H$\alpha$ and $R_{\rm BLRs}\sim295$ light-days is the distance of broad line emission regions to central 
BH estimated through the R-L relation \citep{bd13} with the continuum luminosity at 5100\AA~ about 
$4.94~\times~10^{45}~{\rm erg/s}$. Meanwhile, through properties of total broad H$\alpha$, estimated 
virial BH mass through the formula discussed in \citet{gh05} should be 
\begin{equation}
\frac{M_{\rm BH}}{\rm M_\odot}~\sim~2.2\times10^6\times(\frac{L_{\rm H\alpha}}
	{\rm 10^{42}erg/s})^{0.56}\times(\frac{FWHM_{\rm H\alpha}}
	{\rm 1000km/s})^{2.06}~\sim~1.67\times10^8
\end{equation}
with accepted $L_{\rm H\alpha}\sim(2.29\pm0.11)\times10^{43}{\rm erg/s}$ and line width
$FWHM_{\rm H\alpha}\sim3490{\rm km/s}$ of total broad H$\alpha$, leading the estimated virial BH mass 
to be about one magnitude smaller than the virial BH mass estimated through continuum luminosity and 
second moment of broad H$\alpha$. Here, it is hard to confirm which virial BH mass is preferred in \obj. 
Moreover, besides the two virial BH masses, the BH mass can be also estimated by continuum luminosity  
shown in equation 8 (equation 9\ in \citet{pe04}) 
\begin{equation}
\begin{split}
M_{\rm BH}\sim7.58\times10^7{\rm M_\odot}\times(\frac{\lambda~L_{5100}}{10^{44}{\rm erg/s}})^{0.79}
	~\sim~1.6\times10^9{\rm M_\odot}
\end{split}
\end{equation}
which is roughly consistent with the mean BH mass of the two virial BH masses. Therefore, the BH 
mass $16\times10^8{\rm M_\odot}$ estimated by the continuum luminosity is accepted as the 
BH mass of \obj~ in the manuscript.

%%%%4
	Then, as discussed in \citet{eh95} and in \citet{st03}, the expected disk precession period 
can be estimated as 
\begin{equation}
T_{\rm pre}\sim1040M_{8}R_{3}^{2.5}yr
\end{equation},
where $R_{3}$ and $M_{8}$ mean the distance of optical emission regions to central BH 
in unit of $10^3R_{G}$ ($R_G=\frac{GM_{BH}}{c^2}$) and the BH mass in unit of 
$10^8{\rm M_\odot}$. Considering periodicity about 3.8yr, the expected $R_3$ could be 
around 0.035\ in \obj. However, based on the discussed distance of NUV emission regions 
to central BHs in \citet{mc10} through the microlensing variability properties of eleven 
gravitationally lensed quasars, the NUV 2500\AA~ continuum emission regions in \obj~ 
have distance from central BH as 
\begin{equation}
\log{\frac{R_{2500}}{cm}}=15.78+0.80\log(\frac{M_{BH}}{10^9M_\odot})
\end{equation}
leading the estimated NUV emission regions have distances to central BH about 
$R_{NUV,~3}\sim0.037$. The estimated NUV emission regions have similar distances as the 
optical continuum emission regions in \obj~ under the disk precession assumption, 
strongly indicating that the disk precessions of emission regions are not preferred 
to be applied to explain the detected optical QPOs in \obj.

%%%5
	Moreover, as discussed in introduction, long-term QPOs can be detected in 
blazars due to jet precessions. However, \obj~ is covered in Faint Images of the 
Radio Sky at Twenty-cm (FIRST) \citep{bw95, hw15} 
(\url{http://sundog.stsci.edu/cgi-bin/searchfirst}), but no apparent radio emissions. 
Therefore, jet precessions can be well ruled out to explain the optical QPOs in \obj.

\section{Summaries and Conclusions}
    The final summaries and main conclusions are as follows. 
\begin{itemize}
\item The combined long-term light curve from the CSS and the ZTF can be well described 
by a sinusoidal function with a periodicity about 1400days (3.8yr) in \obj~ even without 
considerations of magnitude difference between CSS light curve and ZTF light curves, 
which can be further confirmed by the corresponding sine-like phase folded light curve.
\item The periodicity can be re-confirmed by the Generalized Lomb-Scargle periodogram 
with confidence level higher than 5$\sigma$, and by the WWZ technique, even without 
considerations of magnitude difference between CSS light curve and ZTF light curves.
\item Moreover, the Pan-STARRS light curves of \obj~ can well follow the sinusoidal 
function described best fitting results to the CSS and ZTF light curves, to provide 
further evidence to support the optical QPOs in \obj. 
\item Under the assumption of a central BBH system in \obj, the BBH system expected 
space separation should be smaller than 0.018pc, based on the upper limit of total BH 
masses $3.3\times10^9{\rm M_\odot}$ estimated through the correlation between BH mass and 
continuum luminosity.
\item Based on the estimated sizes about $37{\rm R_G}$ of the NUV emission regions similar 
as the disk precession expected sizes about $35{\rm R_G}$ of the optical emission regions, 
the disk precession can be not preferred to explain the detected QPOs in \obj. 
\item There are no apparent radio emissions in \obj, strongly supporting that the 
jet precession can be totally ruled out to explain the detected QPOs in \obj.
\item Based on the mathematical CAR process simulating light curves related to the 
intrinsic AGN variabilities, 0.1\% probability can be determined to detect mis-detected 
QPOs in the CAR process simulating light curves, to re-confirm the optical QPOs in \obj~ 
with significance level higher than 3$\sigma$. 
\end{itemize}

\section*{Acknowledgements}
%Zhang gratefully acknowledge the anonymous referee for reading our manuscript carefully and patiently.
Zhang gratefully acknowledge the anonymous referee for giving us constructive comments and 
suggestions to greatly improve our paper. Zhang gratefully acknowledges the kind grant support 
from NSFC-12173020. This paper has made use of the data from the SDSS projects, 
\url{http://www.sdss3.org/}, managed by the Astrophysical Research Consortium for the 
Participating Institutions of the SDSS-III Collaboration. This paper has made use of the data 
from the CSS \url{http://nesssi.cacr.caltech.edu/DataRelease/} and the ZTF 
\url{https://www.ztf.caltech.edu}, and the data from Pan-STARRS \url{https://panstarrs.ifa.hawaii.edu/}, 
and use of the data from the FIRST \url{https://sundog.stsci.edu}. The paper has made use of the public 
JAVELIN code \url{(http://www.astronomy.ohio-state.edu/~yingzu/codes.html#javelin}), and the MPFIT 
package \url{https://pages.physics.wisc.edu/~craigm/idl/cmpfit.html}, and the emcee package 
\url{https://emcee.readthedocs.io/en/stable/}. This research has made use of the NASA/IPAC 
Extragalactic Database (NED, \url{http://ned.ipac.caltech.edu}) which is operated by the California 
Institute of Technology, under contract with the National Aeronautics and Space Administration.

\section*{Data Availability}
The data underlying this article will be shared on reasonable request to the corresponding author
(\href{mailto:aexueguang@qq.com}{aexueguang@qq.com}).

\label{lastpage}
\end{document}